\documentstyle[emulateapj,onecolfloat]{article}

\input epsf
\lefthead{Metzler, White \& Loken}
\righthead{The Effect of the Cosmic Web on Cluster Weak Lensing Mass Estimates}
 
\begin{document}
\twocolumn[
\title {The Effect of the Cosmic Web on Cluster Weak Lensing Mass Estimates}
\author{Christopher A. Metzler$^{1,3}$,
Martin White$^{1,4}$ and
Chris Loken$^{2,5}$}
\affil{$^{1}$Harvard Smithsonian Center for Astrophysics,
60 Garden St, MS-51, Cambridge, MA 02138}
\affil{$^{2}$Department of Physics and Astronomy, St. Mary's University,
Halifax, Nova Scotia, CANADA}
\affil{$^{3}$cmetzler@cfa.harvard.edu}
\affil{$^{4}$mwhite@cfa.harvard.edu}
\affil{$^{5}$cloken@ap.stmarys.ca}

\begin{abstract}
\noindent
\rightskip=0pt
In modern hierarchical theories of structure formation, rich clusters of
galaxies form at the vertices of a weblike distribution of matter, with
filaments emanating from them to large distances and with smaller objects
forming and draining in along these filaments.  The amount of mass
contained in structure near the cluster can be comparable to the
collapsed mass of the cluster itself.  As the lensing kernel is
quite broad along the line of sight around cluster lenses with typical
redshifts $z_{l}\,=\,0.5$, structures many megaparsecs away from the
cluster are essentially at the same location as the cluster itself,
when considering their effect on the cluster's weak lensing signal.
We use large--scale numerical simulations of structure formation in a
$\Lambda$--dominated cold dark matter model to quantify the effect
that large--scale structure near clusters has upon the cluster masses
deduced from weak lensing analysis.  A correction for the scatter in
possible observed lensing masses should be included when interpreting
mass functions from weak lensing surveys.
\end{abstract}

\keywords{cosmology: theory --- galaxies: clusters: general ---
gravitational lensing}

]

\rightskip=0pt

\section{Introduction}

The masses of clusters of galaxies are now measurable by a variety of
observational techniques.  Most approaches require some equilibrium
assumption which relates the shape of the cluster potential to the
energy content of some cluster component.  For example, measuring
or mapping the
temperature of the hot, X--ray emitting intracluster plasma, or
the velocities of cluster galaxies, permits a prediction for the
mass distribution of the cluster.

Another method for estimating cluster masses has been through
observations of weak gravitational lensing of the background galaxy
field by the cluster.  The map of induced distortions in background
galaxy ellipticities can in principle be inverted to provide a
weighted sum of the mass density along the line of sight.  The
weighting is weakest near the lensed sources and near us, while
it is strongest at intermediate redshifts where the cluster lens is
typically located.  Thus, this yields an estimate of the surface
mass density distribution of the cluster and its surroundings, from
which the cluster mass can be inferred.  Since assumptions about
the dynamical or thermodynamic state of the cluster components
are of uncertain validity, while weak lensing analyses probe the
mass distribution directly, estimating cluster masses through
weak lensing analyses has grown extremely popular in the last
decade.  Several groups now have moderate samples of weak lensing
masses (see e.g. Mellier~\cite{Mellier} for a recent review),
while others have applied weak lensing mass estimates to studies
of evolution in the cluster abundance (e.g. Bahcall \&
Fan~\cite{BF}).

One interesting outcome of multi--wavelength studies of clusters
has been that weak lensing mass estimates for clusters sometimes
exceed mass estimates derived from other sources, typically X--ray
observations (see e.g. Squires {\em et.al.}~\cite{A2218}).
This discrepancy is often considered to be
{\em underestimated}, because methods for extracting the mass
from weak lensing data typically depend on the estimated surface
density relative to some value near the edge of the observing
field, which may contain part of the cluster if the viewing
field is small.  When this discrepancy occurs, it is typically
attributed to either the poor quality of the X--ray data
involved --- as good X--ray spectra and images become more difficult
to obtain with
increasing redshift --- or to the failure of equilibrium
assumptions about the state of cluster gas at high--redshift.

However, attempts to reconstruct the mass distribution of clusters
from weak lensing observations are not without pitfalls (see e.g.
Mellier~\cite{Mellier}).  The most well--known of these is
associated with the typically uncertain redshift distribution of the
lensed sources; still others relate to details of the procedure
adopted to go from the observed ellipticity distribution to the mass,
or from instrumental effects.  We here consider another potential
issue:  the effect on mass estimates from clustered matter near
the cluster and in the observing field.

In modern hierarchical models
of structure formation, clusters form in overdense regions at the
vertices of filamentary structures which extend to large distances
from the cluster; they accrete additional mass and smaller collapsed
objects that drain along these filaments.  It is thus reasonable
to expect a beaded filamentary structure surrounding most clusters
of galaxies.  Such overdense filamentary structure, when viewed
in projection through its lensing effects,
could add to the lensing signal produced by a cluster
and result in an overestimate of the cluster mass; in fact, such
an effect may have been identified in one cluster
(Czoske {\em et.al.}~\cite{Czoske}).  In principle clusters could also be
located near voids, leading to a deficit of material along the line of sight
compared to the mean density.
Tentative observational evidence of filamentary structure near clusters
has been reported recently (Scharf {\em et.al.}~\cite{Scharf},
Kull \& Boehringer~\cite{KulBoe}, Kaiser {\em et.al.}~\cite{MS0302}).
A filament lying near or intersecting with the line of sight will
also lens the background
galaxies, and therefore contribute spuriously to the lensing signal.
If the observed lensing signal were attributed solely to the cluster,
the inferred cluster mass could be much larger than its actual mass.

The impact of projection effects upon mass estimators has been
studied in a variety of contexts (see {\em e.g.}
van Haarlem {\em et.al.}~\cite{vH}).
The contamination of weak lensing mass estimates by nearby large--scale
structure has been considered to varying degrees in other
papers (see e.g.~Miralda-Escude~\cite{ME}; Cen~\cite{Cen}; Wambsganss,
Cen \& Ostriker~\cite{WamCenOst}; Reblinsky \& Bartelmann~\cite{RebBart}).
In a recent {\it Letter\/} (Metzler et al.~\cite{MWLN}) we performed
a preliminary study of this effect on three simulated clusters.  We
now broaden this work to consider more clusters and apply a more
accurate modelling of the lensing signal produced by the
simulated clusters.  In \S \ref{sec:mests}, we examine how nearby
large--scale structure affects mass estimates at mean interior
density contrasts
of 200 at a redshift of $z_{l}\,=\,0.5$; we also consider how this
effect depends on cluster redshift and the density contrast within
which masses are measured.  In real situations, however, an
observer is not concerned with the likelihood of finding a certain
lensing mass given a value of the actual cluster mass; instead,
what is desired is the likelihood distribution of a cluster's
{\em actual} mass, given an observation of the lensing mass.  We
consider this in \S \ref{sec:likelihood}.  In \S \ref{sec:massfctn},
we consider the effect of this dispersion in possible observed
lensing masses for a given actual mass upon the observed mass
function of galaxy clusters.  The possibility of avoiding this
source of error by using line of sight velocity histograms to
reject clusters with apparent foreground/background structure
will be addressed in \S \ref{ssec:losvels}.  Finally, we examine
the contribution to this effect by material at successively
larger distances from the cluster in \S \ref{ssec:scaledep}.

\section{Lensing Theory}

In the thin lens approximation, the convergence $\kappa$ is related
to the surface density $\Sigma$ of the gravitational lens by
\begin{equation}
  \kappa = {\Sigma\over\Sigma_{\rm crit}}
         = {4\pi G\over c^2} {D_{A,L} D_{A,LS}\over D_{A,S}}\ \Sigma
\end{equation}
where $D_{A,L}$, $D_{A,S}$, and $D_{A,LS}$ refer to the angular--diameter
distances to the lens and the lensed source, and the angular--diameter
separation between lens and source, respectively.  The convergence
is related to the 2D lensing potential $\psi$
by
\begin{equation}
\label{eq:poisson}
 \kappa\,=\,\frac{1}{2}\nabla^2_{\theta}\psi ,
\end{equation}
with the lensing
potential derived from the peculiar potential $\phi$ induced by
mass inhomogeneities by
\begin{equation}
\psi\,=\,\frac{2}{c^2}\int dD_{L} \frac{\phi}{a}\frac{D_{LS}}{D_{L}D_{S}}
\end{equation}
where here the distances $D_L$, $D_S$, and $D_{LS}$ are {\em comoving},
$a$ is the cosmic scale factor (scaled to 1 at present),
and the integration is taken along the path travelled by the light ray.
With some algebra, and the use of the first Friedmann equation and the
Poisson equation for the peculiar potential $\phi$,
the convergence $\kappa$ can therefore be written as
\begin{equation}
\kappa\, =\, {3\over 2}\Omega_{\rm m}\left(\frac{H_0}{c}\right)^2
           \int dD_L {D_L D_{LS}\over D_S} {\delta\over a},
\end{equation}
where $\delta$ is the local overdensity in terms of the average
density $\bar{\rho}$,
$\delta\,=\,\left(\rho - \bar{\rho}\right)/\bar{\rho}$.
For the specific case of flat universes, $D_{LS}\,=\,D_{S}-D_{L}$,
and so we can write
\begin{equation}
\label{eq:kappaint}
\kappa\,=\, {3\over 2}\Omega_{\rm m}\left(\frac{H_0 D_{S}}{c}\right)^2
           \int dt\,\ t\left(1-t\right) {\delta\over a},
\end{equation}
where $t\,=\,D_L/D_S$.  Note that this derivation for the convergence
along a particular line of sight assumes the lensed source(s) to be
at a single redshift.  For a distribution of sources one further
integrates over $\int dz_s n(z_s)$ where $\int dn=1$.


This last equation, Eq.~(\ref{eq:kappaint}), is the main equation used
in analyzing the simulation datasets.  The important point is that
the integrand can be thought of as the product of the overdensity
and a lensing kernel.  The width of the kernel, written here as
$t\,\left(1-t\right)$, along the line of sight typically does not
vary strongly for clusters at intermediate redshifts, even at large
distances from the cluster lens.  For
instance, for a lens at $z_{l}\,=\,0.5$ and $z_{s}\,=\,1.0$, over
a distance of $256 h^{-1}$Mpc comoving centered on the cluster
(the lengths of the lines of sight we will simulate),
the lensing kernel varies from 96\% to 102\% of its central value
in the flat $\Lambda$--dominated cosmology we are using.  For a lens
at $z_{l}\,=\,1.0$ and $z_{s}\,=\,1.5$, the lensing kernel varies
from 87\% to 111\% of its central value over the same distance.
Therefore, mass concentrations out
to large radii from the cluster can still contribute appreciably
to the lensing signal.

\section{Method}

\subsection{The Cluster Ensemble}

To examine this effect, cosmological simulations including clusters as
well as the large scale structure in which they are embedded are
needed.  Here we have used 12 clusters from the X-Ray Cluster Data
Archive of the Laboratory for Computational Astrophysics of the
National Center for Supercomputing Applications (NCSA), and the
Missouri Astrophysics Research Group of the University of Missouri
(Norman {\em et.al.}\cite{Archive}).
To produce these clusters, a particle-mesh N-body simulation
incorporating adaptive mesh refinement was performed in a volume
$256\,h^{-1}$Mpc on a side.  Regions where clusters formed were
identified; for each cluster, the simulation was then re-run
(including a baryonic fluid) with finer resolution grids centered upon
the cluster of interest.  In the adaptive mesh refinement technique,
the mesh resolution dynamically improves as needed in high-density
regions.  The ``final'' mesh scale at the highest resolution was
$15.6\,h^{-1}$ kpc, with a mean interparticle separation of about
$86\,h^{-1}$ kpc, allowing good resolution of the filamentary
structure around the cluster.  The code itself is described in
detail in Norman \& Bryan (\cite{TheCode}).  Table~\ref{tab:ensstats} shows
the mass within $r_{200}$, the radius containing material at a mean
interior density contrast of 200, at redshifts of $0.5$ and $1.0$
for the 12 clusters in the ensemble.
\begin{table}[ht]
\begin{center}
\caption{
\centerline{Cluster masses within $\bar{\delta}\,=\,200$}
}
\vskip 4.0pt
\label{tab:ensstats}
\begin{tabular}{ccc}
 & \multicolumn{2}{c}{$M_{200}$ ($10^{15}M_{\odot}$)} \\
Cluster & $z\,=\,0.5$ & $z\,=\,1.0$ \\
\hline
0 & 0.83 & 0.24 \\
1 & 0.86 & 0.64 \\
2 & 0.35 & 0.23 \\
3 & 0.58 & 0.25 \\
4 & 0.70 & 0.25 \\
5 & 0.52 & 0.27 \\
6 & 0.52 & 0.38 \\
7 & 0.48 & 0.38 \\
8 & 0.61 & 0.40 \\
9 & 0.64 & 0.26 \\
10 & 0.38 & 0.12 \\
11 & 0.55 & 0.25\\
\end{tabular}
\end{center}
\end{table}

The clusters used here were taken at redshifts of $z\,=\,0.5$
and $z\,=\,1.0$ from simulations of a $\Lambda$CDM model, with parameters
$\Omega_{\rm m}=0.3$, $\Omega_{\rm B}=0.026$, $\Omega_{\Lambda}=0.7$,
$h=0.7$, and $\sigma_8=0.928$.  This dataset comprises
the twelve most massive clusters at
$z\,=\,0$, as determined from the initial, low--resolution run.
Their numbering is in order of $z\,=\,0$ virial mass; as each
cluster evolves, this ranking in mass does not necessarily
hold at higher redshifts.  

\begin{figure}[t]
\begin{center}
\leavevmode
\epsfxsize=6cm \epsfbox{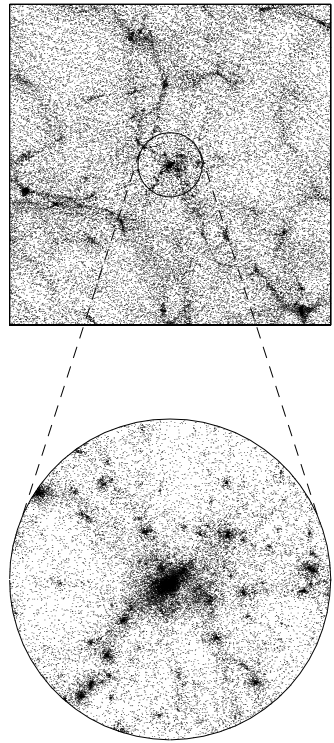}
\end{center}
\figcaption{A slice through the simulation volume, showing simulated
Cluster 6 at $z\,=\,0.5$, at two levels of
magnification.  Dots represent simulation particles.  The slice has
a comoving thickness of $76.8h^{-1}$Mpc.  The top frame shows a window
of width $102.4h^{-1}$Mpc comoving, centered on the cluster; for clarity,
only 1/2 of the mass in the viewing window is shown.   The lower
frame shows a magnification of a $10.2h^{-1}$Mpc radius circle centered
on the cluster; here only 1/8 of the lowest-mass particles are shown for
clarity.
\label{fig:cluster6}
 }
\end{figure}

All of the clusters in the ensemble were embedded in a larger filamentary
network of structure.  The filaments themselves were typically resolved
by the simulation into a string of dense knots embedded in more diffuse
material.  In Fig.~\ref{fig:cluster6}, we show a portion of a slice
through the simulation volume, centered on cluster 6 at $z\,=\,0.5$,
at two successive levels of ``zoom.''  The filamentary structure in
which the cluster is embedded is apparent, despite the limitations of
the two--dimensional image.
Since much of the mass in filaments is at comparatively low density
contrast, the existence of this structure near the line of sight would not
be easy to constrain by observations of redshifts near the cluster.
We will examine this point further in \S \ref{ssec:losvels}

\subsection{Analysis}

For a specific line of sight through a simulated cluster, a map of
the convergence $\kappa$ was constructed by viewing the cluster and
its surroundings through a $40^{\prime}$ square window.  This window was then
divided into a 512 $\times$ 512 grid, and Eq.~(\ref{eq:kappaint}) was
integrated
up along the line of sight through each pixel to produce the final
map.

A potential source of error lies in the integration path used.  Formally,
the integration should be performed along the perturbed path of a
light ray, while we take the integration along the unperturbed path.
However, outside the large density contrasts in the cores of clusters,
we are in the weak lensing regime.  The deflection $\Delta y$
induced by crossing a perturbation over an effective scale $\Delta z$
can be approximated by
\begin{equation}
\Delta y\,\simeq\,\frac{\phi}{c^2}\Delta z,
\end{equation}
where $\phi$ is the magnitude of the potential.  The maximum value
for $\phi/c^2$ to be expected in the simulation volume is that of
the typical value for rich clusters, $10^{-5}$.  This suggests that
the deflection in photon path induced by crossing the simulation
volume is typically smaller than the
filamentary structure of interest, and much smaller than the
cluster radii at a fixed density contrast which we wish to estimate
from the convergence maps.  Furthermore, in real observations,
the measured shear comes from the {\em gradient} in deflection
angle across the image plane, and thus is related to change in
the gradient of the potential across the image plane.  This is
small.  Therefore, the error induced by
integrating straight through the volume, over the unperturbed photon
path, should not appreciably affect the values of the radii
determined from the convergence maps.

Several interesting statistics can be drawn from a convergence
map so obtained.  As an example, multiplying the map by
$\Sigma_{\rm crit}$ for the cluster and lensed source redshifts
of interest transforms the map into a projected surface density
map.  This map can then used to estimate $r_{200}$, the radius
within which the mean interior density contrast is 200.  In three
dimensions, this radius is defined in terms of the enclosed mass by
\begin{equation}
  M\left(<r_{200}\right)=200\times \left({4\pi\over3}\right)
    \Omega_{\rm m}\, \rho_{\rm crit}\, r_{200}^3.
\label{eq:r200def}
\end{equation}
A projected estimate of $r_{200}$ is then extracted from the surface
density map by considering the radius of the circle, centered on the
cluster, which contained the amount of mass given by
Eq.~(\ref{eq:r200def}) above, i.e.
\begin{equation}
\label{eq:projestdef}
 \int_0^{2\pi}{\rm d}\theta \int_0^{r_{200}} R\, {\rm d}R
\,\ \Sigma\left(R,\theta\right)
   = M\left(<r_{200}\right)
\end{equation}
with $\Sigma\left(R,\theta\right)$ the surface density on the map
in terms of a two-dimensional radius $R$.  An estimate of the
radius at a density contrast of 500 can be obtained by a similar
procedure.  Note that this approach implicitly assumes that all
the mass in the convergence map is associated with the cluster;
except for lines of sight which are substantially underdense
outside the cluster, this approach should result in overestimates
of radii at a fixed overdensity, and thus of the mass at those
radii.  However, the scatter in such estimates, for different
lines of sight, is driven by the dispersion in mass outside the
cluster but inside a line of sight's viewing window.  Unless an
estimator makes an explicit attempt to correct for such contamination,
the scatter in this simple projected estimate should be comparable
to that in a different estimator.


The technique of aperture densitometry allows another useful
quantity to be extracted from the convergence map:  the so--called
$\zeta$ statistic, defined as the mean value of the convergence
$\kappa$ within a circular area on the sky of radius $r_1$
minus the mean value within a bounding annulus
$r_1\le r\le r_2$ (Fahlman et al.~\cite{Fahlman},
Kaiser~\cite{K95}).
\begin{equation}
\label{eq:zetadef}
  \zeta(r_1,r_2)\,=\, \left\langle \kappa( 0 ,r_1) \right\rangle -
                          \left\langle \kappa(r_1,r_2) \right\rangle.
\end{equation}
This quantity can be written as an integral of the tangential shear
$\gamma_t$,
\begin{equation}
\label{eq:zetadefshear}
\zeta(r_1,r_2)\,=\, {2\over 1-r_1^2/r_2^2} \int_{r_1}^{r_2} {dr\over r}
                          \left\langle \gamma_t \right\rangle \qquad ,
\end{equation}
and it is this way that the $\zeta$ statistic is normally measured
from the observational data.  However, it can also be measured
from mock convergence maps constructed from our data.  If the
mean convergence in the outer annulus is thought of as an estimate
of the ``background'' contribution to $\kappa$ everywhere, then
$\zeta$ provides an estimate of the convergence signal which
comes from the cluster alone.  Multiplying $\zeta$ through by
$\Sigma_{\rm crit}$ then gives a surface density, which can be manipulated
as above to find the radius at a density contrast of 200.  This,
in principle, can be thought of as an attempt to correct for the
projection effect we consider here; note that if the second,
subtracted term in Eq.~(\ref{eq:zetadef}) were not present, the
masses found from the $\zeta$ statistic would be identically those
found from Eq.~(\ref{eq:projestdef}).

Before examining each cluster and extracting such statistics, the
particle dataset was cut down to a
$128\,h^{-1}$Mpc radius sphere centered on the cluster of interest.
This guarantees that different lines of sight through the cluster do
not include additional mass simply by geometry, by being near
diagonals of the simulation cube.  Since the radius of the spherical
dataset is very large compared to the radii at a fixed density
contrast obtained for each cluster, no significant radial surface
density gradient is introduced by a decreasing chord length through
the sphere with projected radius.

\begin{figure}[t]
\leavevmode
\epsfysize=8cm \epsfbox{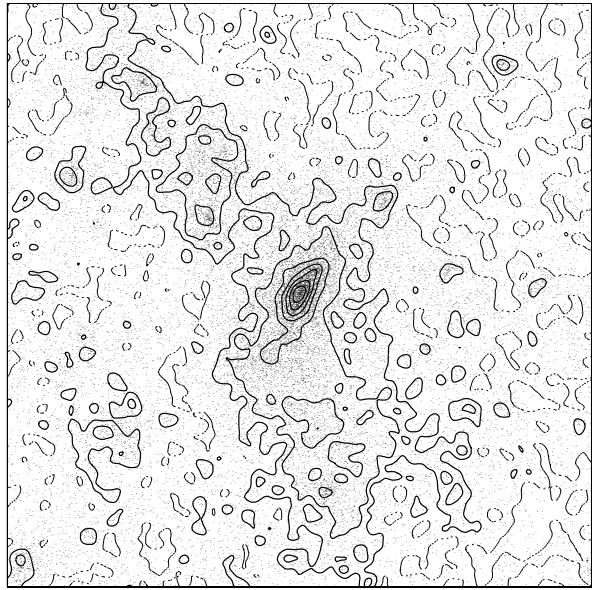}
\figcaption{View of Cluster 6 at $z\,=\,0.5$ in projection through the
simulation volume.  The window is $1600^{\arcsec}$ on a side, corresponding to
a linear size of $10.3h^{-1}$Mpc at $z\,=\,0.5$.  Dots represent simulation
particles; only one-eighth (12.5\%) of the lowest mass particles in the viewing
window are shown.  Contours overlaid on the plot show lines of constant
convergence $\kappa$ from the convergence map determined from the mass
distribution in the viewing window.  The lowest solid contour is at $0.02$,
while all other contours step up by $0.04$ from $\kappa\,=\,0.04$.  The
dotted line marks contours where the calculated convergence is zero,
and thus denotes areas of the map which show a negative value for $\kappa$.
This viewing angle corresponds to a mass estimate within a density
contrast of 200 of
$M_{\rm lens}\,=\,2.7M_{\rm true}$.
\label{fig:examplemap}
}
\end{figure}

We observed each of the 12 clusters used from $5000$ randomly
chosen viewing angles, for each of the two redshifts studied.  For
each cluster and viewing angle, a map of the convergence $\kappa$
was constructed using the formalism described in the previous section.
To illustrate the result of this process, Fig.~\ref{fig:examplemap}
shows a view of Cluster 6 at $z_{l}\,=\,0.5$ along one particular line
of sight, along
with contours showing the convergence map resulting from the procedure
described in the previous section.
We take the lensed sources to be at a redshift
$z_{s}\,=\,1.0$ for the clusters studied at $z_{l}\,=\,0.5$, while
the sources were assumed to be at $z_{s}\,=\,1.5$ when examining
the clusters at $z_{l}\,=\,1.0$.  Note that by placing the lensed
sources at a fixed, known redshift, we are ignoring the potentially
large source of error associated with an unknown source redshift
distribution.

With the convergence map, and thus an implied surface density map,
in hand, a lensing estimate of $r_{200}$ was then obtained by
determining the radius at which the mass given by Eq.~\ref{eq:projestdef}
equals the mass inferred from Eq.~\ref{eq:r200def} --- that is,
the radius at which the interior mass in the surface density map
would be at a density contrast of 200 if contained within a sphere
of that radius.  This radius was compared to the cluster's
true $r_{200}$, extracted from the three--dimensional mass distribution.
The ratio of the projected mass to true mass within a density contrast
of 200 is given simply by the cube
of the ratio of the estimated value of $r_{200}$ to the true, 3D value.
For each cluster, a value of this ratio was obtained for each viewing angle.
This process was repeated $5000$ times for each cluster at each redshift.
The same procedure was followed to generate values of $r_{500}$.  The
probability of encountering a particular ratio of
$M_{\rm lens}/M_{\rm true}$ (where by $M_{\rm true}$ we mean the
actual mass within a 3D radius containing the chosen overdensity)
was then examined by plotting histograms
of the results of this procedure.

In preparing these histograms for analysis, lines of sight producing
ratios greater than 2
(i.e. estimated masses off by more than 100\%) were excluded
from the histograms.  In many cases, such lines of sight pass through
a second rich cluster, of greater mass than the one being studied.
In the real universe, such a situation should be detectable through the
distribution of galaxy redshifts in the viewing field; typically,
the more massive cluster is likely to be the one of interest,
resulting in a ratio for that cluster less than would occur for the
smaller cluster.  Of course, it remains possible that some lines of
\begin{figure*}
\begin{center}
\leavevmode
\epsfysize=18cm \epsfxsize=18cm \epsfbox{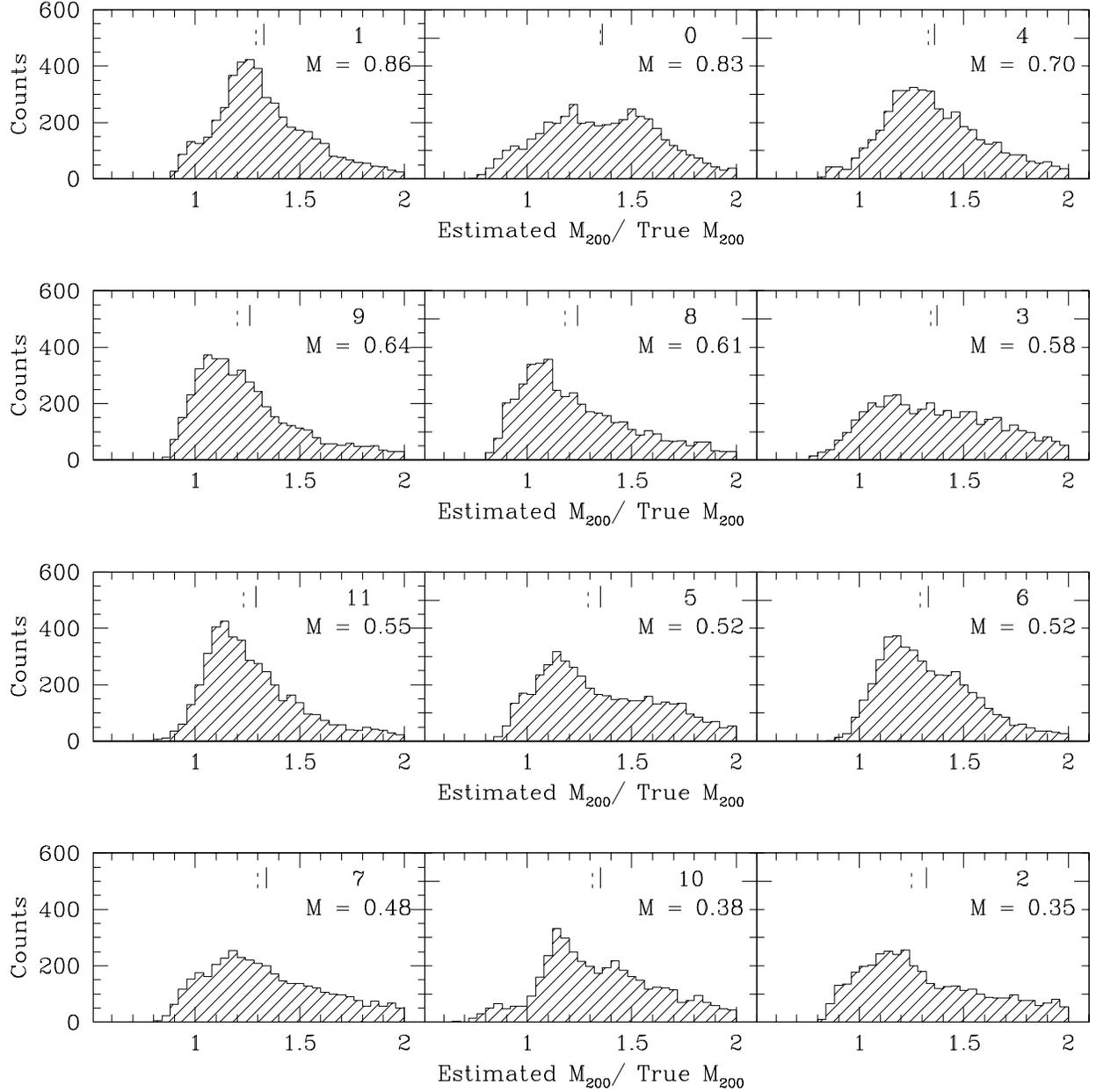}
\end{center}
\figcaption{Histograms of $M_{\rm lens}/M_{\rm true}$, measured within a mean
interior density contrast $\bar{\delta}\,=\,200$, at z=0.5.  Clusters are
ordered by $M_{\rm true}$, with the mass given for each cluster in units
of $10^{15}$ M$_{\odot}$.  Lines of sight near the box principle axes,
through identified large clusters, or producing mass estimates twice
that of the cluster, are all excluded.  The vertical hash mark near
the top of each box indicates the location of the mean for that
cluster's histogram.
\label{fig:hist200z0.5}
}
\end{figure*}

sight with $M_{\rm lens}/M_{\rm true}\,>\,2$ would be produced through
lines of sight which do not pass through a larger cluster, and
thus would not be excluded so easily by the actual observer.  However,
in the spirit of making a conservative estimate of the magnitude of
this effect, we uniformly exclude lines of sight with ratios so
large.  Furthermore, any lines of sight which pass within
$3 h^{-1}$Mpc of another rich cluster in the simulation volume, but
were not caught by the ``factor of 2'' limit just described, were
similarly excluded.

Also excluded were lines of sight which were within $10^{\circ}$ of
one of the principle axes of the simulation volume.  The issue here
is that the simulations incorporated periodic boundary conditions;
along a principle axis of the box, structure at opposite ends of
the line of sight through the simulation volume are correlated.
Therefore,
a density enhancement at one end makes a density enhancement at
the other end more likely.  Excluding lines of sight near the 
simulation volume's principle axes reduces any spurious signal
produced by the periodic boundary conditions.

\section{Effect of Structure On Mass Estimates}
\label{sec:mests}

\subsection{Measurements of $M_{200}$ at $z_{l}\,=\,0.5$}

We begin by considering the effect of the network of structure in
which the cluster is embedded on simple estimates of
the mass within a mean interior density contrast of 200.
As noted
above, the ratio of the estimated lensing mass within this density
contrast to the actual, 3--D mass containing matter at this mean
density contrast, is given by $\left(r_{200,est}/r_{200,3D}\right)^3$,
where $r_{200,est}$ is determined from the convergence map as
described above, and $r_{200,3D}$ is determined directly from
the full 3--D mass distribution.  Fig.~\ref{fig:hist200z0.5}
shows histograms of the results of this for each cluster in the
ensemble.  Here the clusters were observed at a redshift of
$z_{l}\,=\,0.5$; the lensed sources were assumed to lie at
a redshift $z_{s}\,=\,1.0$.  The histograms show the resulting
ratio of estimated--to--actual mass for the $5000$ random viewing
angles used.    The clusters are ordered in the plot by their mass
at this redshift, from heaviest to lightest.

Several general properties of the histograms in Fig.~\ref{fig:hist200z0.5}
are worth noting.  First, the occasional ``spiky'' nature of the histograms
does not come from shot noise; instead, it is due to discrete objects
being inside or outside the visual field.  As an example, a small halo of
matter near the cluster will project entirely within the actual 3D
$r_{200}$ for a fraction of the lines of sight.  For any such line of
sight, the effect on the estimated value of $r_{200}$ is identical.
Second, the histograms are strongly positively skewed, even after
excluding lines of sight that are likely to generate large positive
biases in the estimated mass.

The most important point to note from these histograms, however, is
the magnitude of the dispersion in possible values of the mass ratio.
The large dispersion is {\em not} induced by
the anisotropic structure of the cluster itself; this was checked by
regenerating the histogram for one of the clusters using a subset of
the simulation particles intended to represent the cluster alone.
This was done by identifying particles located at and around the
cluster at local density contrasts above $\delta\,=\,70$ (chosen
because density profiles near $r^{-2}$ reach a local density contrast
near 70 at a mean interior density contrast of 200).  This set was
identified as the cluster, and a small sphere containing this subset
but little nearby material was then cut out of the simulation volume.
The histogram produced by viewing the clearly prolate cluster at a
large number of randomly chosen viewing angles produced a much
narrower distribution, with a maximum offset of less than $10\%$ in
the mass ratio and a mean offset of approximately half that value.

We will see in \S \ref{ssec:scaledep} that the mean values
of the histograms are driven by material within 20 Mpc of the
cluster; material outside this distance serves primarily to widen the
dispersion in possible values of $M_{\rm lens}/M_{\rm true}$ resulting from
an observation.  While the mass estimator used herein is simplistic,
and by construction was expected to produce a positive
bias in the mass estimate, any estimator which might correct for such
bias (by, for instance, assuming a model for the radial distribution
of matter outside clusters) will still run afoul of this large
dispersion.  Large discrepancies between the weak lensing mass and
the virial mass of clusters are possible.

\begin{figure}[ht]
\leavevmode
\epsfysize=8cm \epsfbox{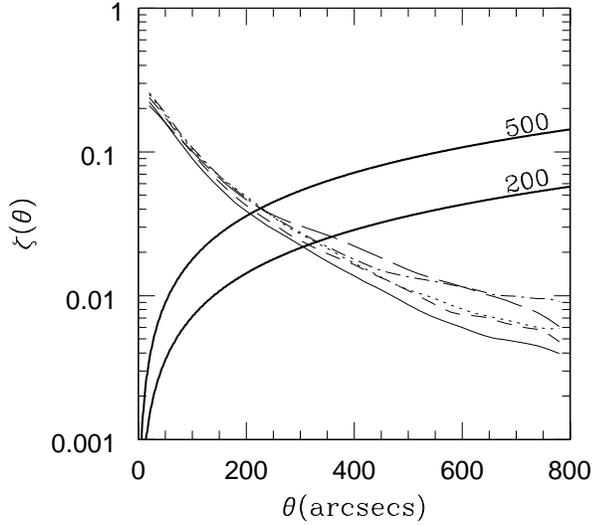}
\figcaption{Aperture densitometry plot for one cluster (6), showing
$\zeta$--profiles for five lines of sight, as well as curves which mark the
value of $\zeta$ at a given radius for a mean interior density contrast
of 200 or 500.  Here $\theta$ defines the inner radius of the aperture
used for each point, with the outer radius at 800''.
The lines of sight used were chosen because the simple
projected estimator, applied to each line of sight, returned a value
of 1.00 (solid), 1.25 (dotted), 1.50 (short dashed), 1.75 (long dashed),
and 2.00 (dot--dashed).
\label{fig:zetacl06z0.5}
}
\end{figure}

To illustrate using another mass estimator,
in Fig.~\ref{fig:zetacl06z0.5}, we show aperture densitometry
plots for Cluster 6, for five lines of sight through the cluster.
The lines of sight were chosen to span a range in $M_{\rm lens}/M_{\rm true}$
from the simple projected estimator of from $1.0$ to $2.0$.
We have taken an outer radius of $\theta_2=800''$, within the
half--degree field of view typical of new large CCD cameras.
We have explicitly checked that reducing the radius to half this
value does not change our result.
Also shown are two curves marking the value required of $\zeta$ at
a given radius for that radius to enclose a given estimated
density contrast.  For example, where a particular $\zeta$ profile
crosses the line labelled ``200'' marks the radius that aperture
densitometry would suggest is $r_{200}$ for that line of sight.
If a $\zeta$ profile lies above that curve, the mean interior
density contrast at that radius is higher than 200, while below
implies lower than 200.  Amongst the lines of sight shown, the
largest estimated $r_{200}$ is a factor of $1.18$ larger than the
smallest, corresponding to a factor of $1.63$ difference in mass;
this should be compared to the factor of two difference between
largest and smallest masses predicted by the simple projected
estimator.  Thus, the masses predicted from aperture densitometry
appear to
have a dispersion tighter than given by simple projected mass
estimates, but still
quite substantial; such techniques as aperture densitometry can
ameliorate our problem, but do not resolve it.
None of the curves shown clearly
indicate anything untoward about the line of sight shown; an
observer would not be driven to suspect a strong bias in the
estimated mass from the shape of the $\zeta$ profile even in
the most extreme cases shown here.

\begin{figure}[b]
\leavevmode
\epsfysize=8cm \epsfbox{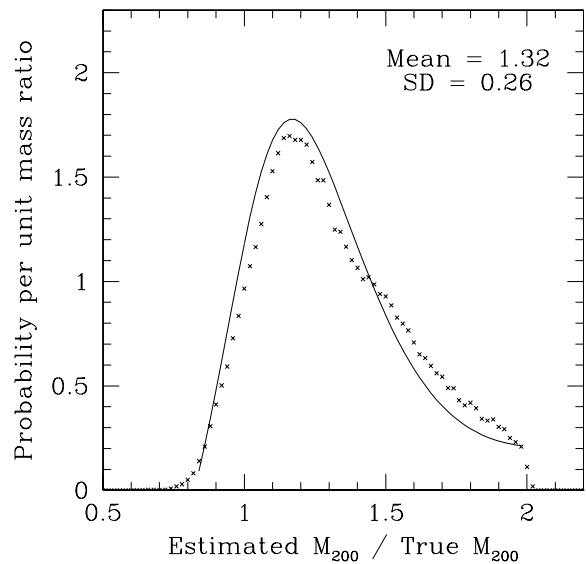}
\figcaption{A PDF of $M_{\rm lens}/M_{\rm true}$ for the ensemble at $z\,=\,0.5$.  The points mark the PDF as derived from the ensemble, while the solid
curve shows an intentionally conservative polynomial approximation to
the data.
\label{fig:pdfz0.5}
}

\end{figure}

The histograms in Fig.~\ref{fig:hist200z0.5} were used to construct
a cumulative probability distribution function for the possible values
of $M_{\rm lens}/M_{\rm true}$.  A plot of the PDF is shown in
Fig.~\ref{fig:pdfz0.5}.  Also shown is a simple approximation to
the shape of the PDF with a smooth curve.  As there is no theoretical
prejudice in favor of any particular shape for the curve used,
a simple approximation using polynomials was constructed
by hand, for use later in \S \ref{sec:likelihood}.  It errs
on the conservative side, in that the strength of the effect predicted
by the smooth curves is less than is actually seen in the simulated
data; the mean of the curve shown is 1.28 with a dispersion of 0.23,
in contrast to the actual data which show a mean of 1.32 and a
dispersion of 0.26.
This is again in the spirit of suggesting a conservative lower bound
for the size of the effect in \S \ref{sec:likelihood}.

\subsection{Measuring masses at higher density contrast}

\begin{figure*}
\begin{center}
\leavevmode
\epsfysize=18cm \epsfxsize=18cm \epsfbox{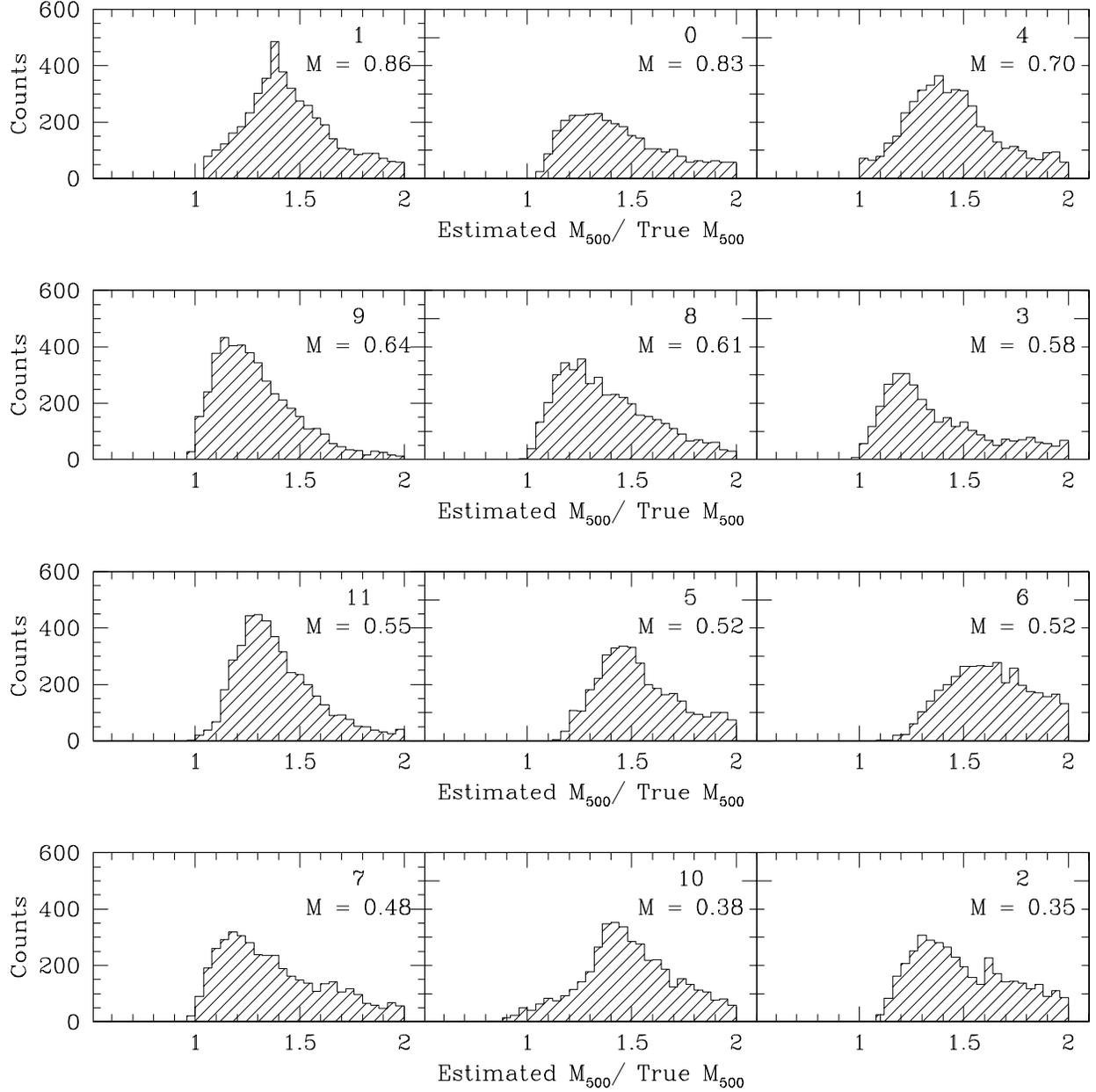}
\end{center}
\figcaption{Histograms of $M_{\rm lens}/M_{\rm true}$, measured within a mean
interior density contrast $\bar{\delta}\,=\,500$, at z=0.5.  Clusters are
ordered by $M_{\rm true}$, with the mass given for each cluster in units
of $10^{15}$ M$_{\odot}$.
\label{fig:hist500z0.5}
}
\end{figure*}

Note that in Fig.~\ref{fig:zetacl06z0.5}, the dispersion in estimated
values of $r_{500}$ is much tighter for the lines of sight shown.  This
suggests that measuring masses within a higher mean interior density
contrast reduces the magnitude of this effect.  Fig.~\ref{fig:hist500z0.5}
repeats the exercise in constructing Fig.~\ref{fig:hist200z0.5},
but with measuring the estimated and actual mass within a density
contrast of $500$.

From these histograms, it does not appear that any such reduction
in the magnitude of the dispersion occurs.  The mean for the cumulative
PDF of these histograms is $1.44$, much larger than the mean of
$1.32$ found for the $r_{200}$ histograms; the standard deviation
is $0.23$, only slightly (but statistically significantly) smaller
than for the $r_{200}$ case.  There appears to be little reduction
in dispersion, and no reduction in bias, by going to $r_{500}$ with
the simple projected estimator.  The reason behind not seeing
any reduction in the bias here, yet apparently seeing such a
reduction in the predictions of the lines of sight examined in
Fig.~\ref{fig:zetacl06z0.5},
is likely attributable to the effect of subtracting the outer annular
mean of the convergence in constructing $\zeta$.  Without this term
in the definition of $\zeta$ --- if $\zeta$ were defined solely
by the average $\kappa$ within a given radius --- then the masses
thus calculated would be simply a product of the surface area within
that radius and $\zeta$, the average value of the convergence.  This
is identically the same process as is used to find Fig.~\ref{fig:zetacl06z0.5}.
Therefore, the decrease in dispersion must be from subtracting off
the outer annulus.  In other words, whether measuring masses at
a higher density contrast reduces the magnitude of the dispersion
appears to depend on the estimator used; determining the dispersion
of a proposed estimator through methods such as used here is important
for understanding the results of the estimator.

Previously, studies of the accuracy of cluster X--ray binding mass
estimates showed that such estimates were much more robust when
measured within a mean interior density contrast of
$\bar{\delta}\,=\,500$ than when
$\bar{\delta}\,=\,200$ was used, for reasons related to the dynamical and
thermodynamic state of the intracluster gas (Evrard, Metzler \&
Navarro~\cite{EMN}).  It is somewhat surprising that no comparable
result appears here.  Since the means of the histograms displayed
in Fig.~\ref{fig:hist200z0.5} show range from $1.24$ to $1.37$,
it is quite possible that sample variance plays a role here.

\subsection{Evolution with redshift}

Also of interest is how the evolution of structure affects the
magnitude of this effect.  We might expect the contamination by
projected structure to be {\em worse} for clusters at higher
redshift.  Consider a cluster at $z\,=\,0.5$ from a particular
line of sight.  As the cluster evolved from $z\,=\,1.0$ to
$z\,=\,0.5$, the amount of projected mass within a $40^{\prime}$
field of view likely changed little.  What changed, instead,
was the amount of this mass contained within the cluster of
interest, and the amount contained in smaller halos and nearby
filamentary structure that merged with the cluster in the
intervening time.  Thus, large clusters with a small amount
of foreground and background material (and thus a small effect
on the mass estimate) at $z\,=\,0.5$ were likely smaller
clusters with a somewhat larger amount of foreground and background
material (and thus a larger effect on the mass estimate) at
$z\,=\,1.0$.
\begin{figure*}
\begin{center}
\leavevmode
\epsfysize=18cm \epsfxsize=18cm \epsfbox{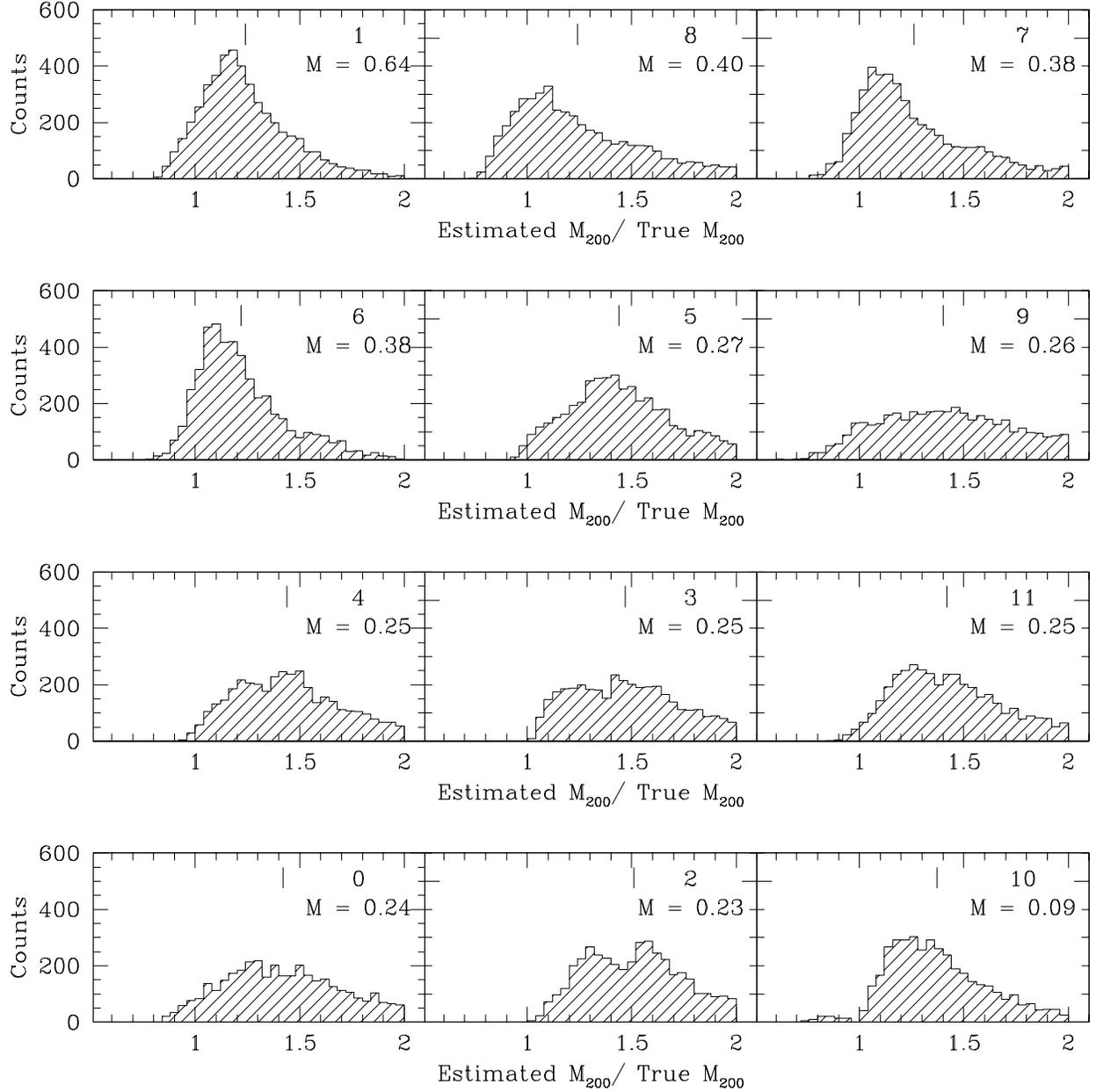}
\end{center}
\figcaption{
Histograms of $M_{\rm lens}/M_{\rm true}$, measured within a mean
interior density contrast $\bar{\delta}\,=\,200$, at z=1.0.  Clusters are
ordered by $M_{\rm true}$, with the mass given for each cluster in units
of $10^{15}$ M$_{\odot}$.  Lines of sight near the box principle axes,
through identified large clusters, or producing mass estimates twice
that of the cluster, are all excluded.  The vertical hash mark near
the top of each box indicates the location of the mean for that
cluster's histogram.
\label{fig:hist200z1.0}
}
\end{figure*}

Fig.~\ref{fig:hist200z1.0} shows the $M_{\rm lens}/M_{\rm true}$ histograms
for $M_{200}$ estimated for the clusters at $z\,=\,1.0$.  Again, the
clusters are ordered by their mass at that redshift, with the mass
in units of $10^{15}M_{\odot}$ shown, and a vertical hash mark denoting
the mean of each histogram.  By eye, for most clusters, both the means
and the dispersions are consistently larger than their counterparts
at $z\,=\,0.5$.  This is not always true; Cluster 6 is an exception
\begin{figure*}[ht]
\begin{center}
\leavevmode
\epsfysize=8cm \epsfxsize=16cm \epsfbox{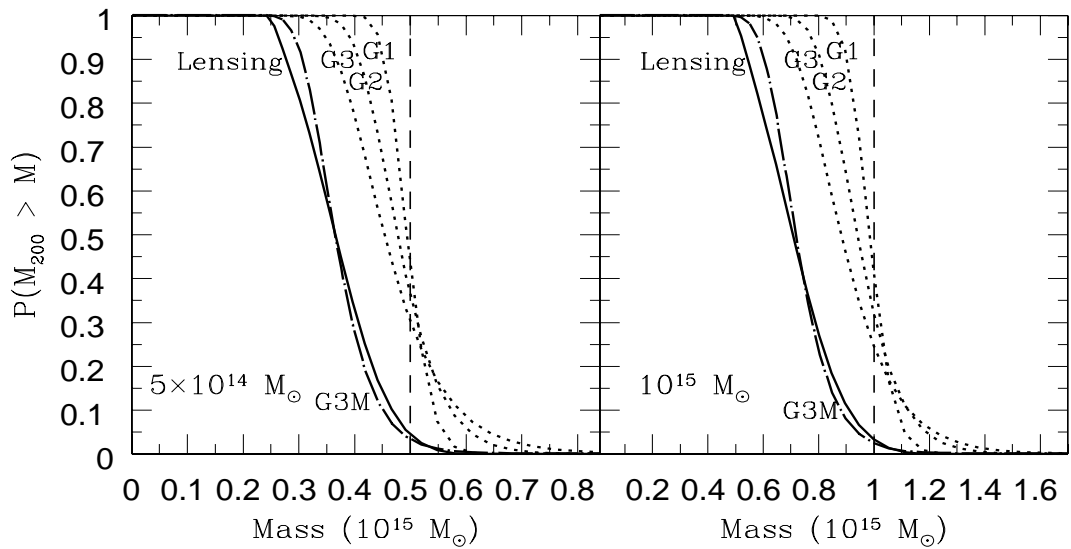}
\end{center}
\figcaption{The likelihood that a cluster with an observed lensing mass
within $\bar{\delta}\,=\,200$ of $M_{\rm lens}\,=\,5\times 10^{14}M_{\odot}$
(left) or $M_{\rm lens}\,=\,10^{15}M_{\odot}$ (right) has
an actual mass within $\bar{\delta}\,=\,200$ greater than or equal
to $M$, at $z\,=\,0.5$.  The observed $M_{\rm lens}$ is marked by a dashed
line.  Considered are six possible PDFs mapping the probability of
finding a particular observed mass given an actual mass.  The heavy
solid line to the left in each case shows the likelihood curve for the
PDF we derived in Fig.~\ref{fig:pdfz0.5}.  The three dotted curves
marked G1, G2, and G3 show the result for Gaussian PDFs in the mass
ratio centered on $M_{\rm lens}/M_{\rm true}\,=\,1$, with dispersions of
$0.1$, $0.2$, and $0.3$ respectively.  The dot-dashed curve marked
$G3M$ shows the result for a Gaussian PDF with a dispersion of $0.3$,
but with the mean ratio shifted to $1.3$, comparable to that of the
PDF in Fig.~\ref{fig:pdfz0.5}.
\label{fig:likelihood}
}
\end{figure*}
to this rule.  A coadded histogram of the $z\,=\,1.0$ data yields a
mean and dispersion of $1.36$ and $0.27$, compared to $1.32$ and $0.26$
at $z\,=\,0.5$.  The effect is not large but, coming from 5000 lines
of sight, is statistically significant; for instance, the difference
between the two means is significantly larger than the uncertainties in
their values.  Thus, while we continue our analysis using the
$z\,=\,0.5$ dataset, to be conservative about the magnitude of this
effect, it should be noted that the problem is likely worse for
clusters at still higher redshift.

\section{Likelihood of Cluster Actual Mass Given An Observed Estimate}
\label{sec:likelihood}

In the previous section, we considered the distribution of possible
values of the ratio $M_{\rm lens}/M_{\rm true}$ --- that is, the distribution
of possible observational estimates for the mass of a cluster given
its true mass within the density contrast of interest.  In the real
universe, however, the problem faced by an observer is the opposite:
given an estimate of the mass taken from observations, what might
the actual mass of the cluster be?  The histograms we examined in
the previous section give us the probability of observing a certain
lensing mass given a certain true mass, $P\left(M_{\rm lens}|M_{\rm true}\right)$;
in real situations, we typically desire the the inverse quantity, the
likelihood of a true mass of a given value, given an observed lensing
mass, $P\left(M_{\rm true}|M_{\rm lens}\right)$.

We can examine this by considering clusters with an observed mass
$M_{\rm lens}$.  If the probability of a cluster with a true mass $M$
being observed with mass $M_{\rm lens}$ is given by
$P\left(M_{\rm lens}|M\right)$, and if the {\em number density} of clusters
with true mass in the range $\left(M, M\,+\,{\rm d}M\right)$ is given
by $n\left(M\right) {\rm d}M$, then the product
$P\left(M_{\rm lens}|M\right)\ n\left(M\right) {\rm d}M$ gives the
number density of clusters with true masses in the range
$\left(M, M\,+\,{\rm d}M\right)$ which are then observed to
have an effective lensing mass $M_{\rm lens}$.  Since the total number
density of clusters with observed mass $M_{\rm lens}$ should be given by
an integral of this function, we can find the likelihood of interest
by forming the fraction.  In other words,
\begin{equation}
\frac
{\int_{M_1}^{M_2} P\left(M_{\rm lens}|M\right) n\left(M\right){\rm d}M}
{\int_{0}^{\infty} P\left(M_{\rm lens}|M\right) n\left(M\right){\rm d}M}
\end{equation}
gives the probability that a cluster with an observed lensing mass
$M_{\rm lens}$ has an actual mass in the range $\left(M_1, M_2\right)$.

We can understand the qualitative nature of the effect by examining
the terms of the integral in the numerator.  In the previous section,
we noted the biased form of the $M_{\rm lens}/M_{\rm true}$ distribution; this
argues that observed masses are likely overestimates of the true mass
of a cluster.  One can imagine that this bias could be corrected for,
using an estimator that takes contamination of the mass estimate by
foreground and background mass into account.  This would remove such
a bias, but not the distribution of the scatter about the mean.

However, we must also consider the effect of the cluster mass function,
$n\left(M\right) {\rm d}M$, the number density of clusters as a function
of mass.  Theory and observations both strongly suggest that this is
a steeply falling function with mass; there are more low mass clusters
than there are high  mass clusters.  Because of this, even if the
distribution of observed lensing masses $P\left(M_{\rm lens}|M_{\rm true}\right)$
from the previous section were symmetric about the mean --- even if
there were no bias in the mass estimator --- it would still be more
likely to overestimate masses than underestimate them.  To explain this
in more concrete terms, consider a cluster with an observed lensing mass
$M_{\rm lens}\,=\,10^{15}M_{\odot}$.  If we consider two values of the
actual cluster mass $M_1\,=\,8.33\times 10^{14}M_{\odot}$ and
$M_2\,=\,1.25\times 10^{15}M_{\odot}$ (mass ratios $M_{\rm lens}/M_{\rm true}$
of $1.2$ and $0.8$ respectively), then even if the probability
that clusters of masses $M_1$ and $M_2$ will be observed at lensing
mass $M_{\rm lens}$ is the same (as would be the case if our histograms
were unbiased and symmetric), it is still more likely that a given
cluster observed at $M_{\rm lens}$ has mass $M_1$ than $M_2$, simply
because there {\em are more clusters} at mass $M_1$ than $M_2$.

Thus, {\em we expect that weak lensing masses for clusters systematically
overestimate the true masses of clusters} within the density contrast
of interest.  To be fair, this effect should occur with any mass estimator,
whether based on lensing, hydrostatic, or dynamical arguments since any
estimator is bound to have some scatter in its predictions.  The magnitude
of the overestimate is dependent on two things:  the width of the
distribution $P\left(M_{\rm lens}|M_{\rm true}\right)$, which for weak lensing
we considered in the previous section; and the steepness of the mass
function at the observed value $M_{\rm lens}$.  The massive clusters
($M_{200}\,\simeq\,10^{15}M_{\odot}$) at moderate redshift that provide
the typical objects for weak lensing analysis lie on or near the
exponential cutoff of the theoretical Press--Schechter mass function;
the magnitude of these overestimates can be expected to be quite strong.

In Fig.~\ref{fig:likelihood}, we show the likelihood that a cluster
at $z\,=\,0.5$, observed to have a lensing mass within a density
contrast of 200 of either $M_{\rm lens}\,=\,5\times 10^{14}M_{\odot}$ or
$10^{15}M_{\odot}$, has an actual
mass within a density contrast of 200 equal to or greater than $M$.
To construct this figure, the theoretical Press--Schechter mass
function for the $\Lambda$CDM model assumed in this paper was used.
Several different curves relating possible values of the observed
mass to the true mass were considered.  For the lensing mass, the
curves shown indicate that it is 100\% likely that the actual cluster
mass is greater than half the observed mass; this merely reflects the
fact that in constructing $P\left(M_{\rm lens}|M_{\rm true}\right)$, we
artificially truncated the error histograms at
$M_{\rm lens}/M_{\rm true}\,=\,2$, as explained earlier.
The important point to take from this plot is that regardless of the
magnitude of the dispersion, it is unlikely that the cluster actually
has the observed mass or greater.  Even without a bias, a dispersion
of $30\%$ in the mass ratios indicate that a cluster observed at
a mass of $5\times 10^{14}M_{\odot}$ is $70\%$ likely to be of
lower mass, while a cluster at $10^{15}$ is over $75\%$ likely
to be of lower mass.  The situation is far
worse if the dispersion is larger or if a bias exists.

\section{The Mass Function}
\label{sec:massfctn}

If there is some dispersion in possible observed masses for clusters
--- that is, if the mapping from actual mass to observed mass is not
one--to--one --- then there will also be an effect on the observed
mass function.  The number density of clusters of some observed
mass $M_{\rm lens}$ will actually be comprised of contributions from
a range of actual cluster masses.  This can be described by a convolution
of the true mass function with the probability of observing a cluster
of mass $M_{\rm true}$ at a given mass $M_{\rm lens}$.  Mathematically,
the abundance of clusters with observed masses in the range
$\left(M_{\rm lens}, M_{\rm lens}+dM_{\rm lens}\right)$ should be given by
\begin{displaymath}
n\left(M_{\rm lens}\right) dM_{\rm lens}\,=\,\hspace{2.2in}
\end{displaymath}
\begin{equation}
dM_{\rm lens}\int^{\infty}_{0}P\left(M_{\rm lens}|M_{\rm true}\right)n\left(M_{\rm true}\right) dM_{\rm true}
\end{equation}
where, as earlier, $P\left((M_{\rm lens}|M_{\rm true}\right)$ is the probability
that a cluster with an actual mass of $M_{\rm true}$ will be observed with
mass $M_{\rm lens}$.
\begin{figure}[b]
\begin{center}
\leavevmode
\epsfysize=8cm \epsfbox{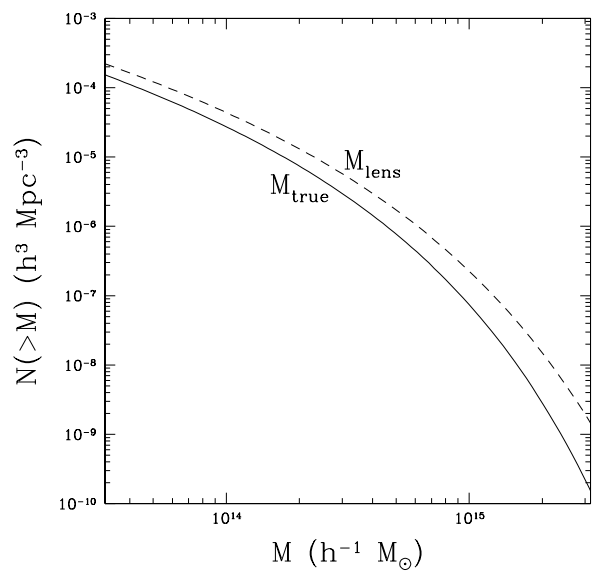}
\end{center}
\figcaption{
The number density of clusters predicted to lie above a given mass
$M$.  The solid line indicates the actual mass function predicted
by Press-Schechter, while the dashed line indicates the abundance
as a function of observed $M_{\rm lens}$.
\label{fig:NaboveM}
}
\end{figure}
Fig.~\ref{fig:NaboveM} demonstrates the result of this convolution,
using the PDF derived for Fig.~\ref{fig:pdfz0.5}.  Here an extra
integral is done, from a given mass to infinity, to find the
number density of clusters above a given mass.  Plotted are both
the expectation for the number density above a given observed mass
value $M_{\rm lens}$ (dashed line) and the number density above a
given actual mass $M_{\rm true}$.  The figure suggests that lensing
observations would indicate three times as many clusters at a mass
of $10^{15}M_{\odot}$ as are actually present.  This is significant,
but certainly insufficient to mimic the lack of strong evolution
\begin{figure*}[t]
\begin{center}
\leavevmode
\epsfysize=14cm \epsfxsize=16cm \epsfbox{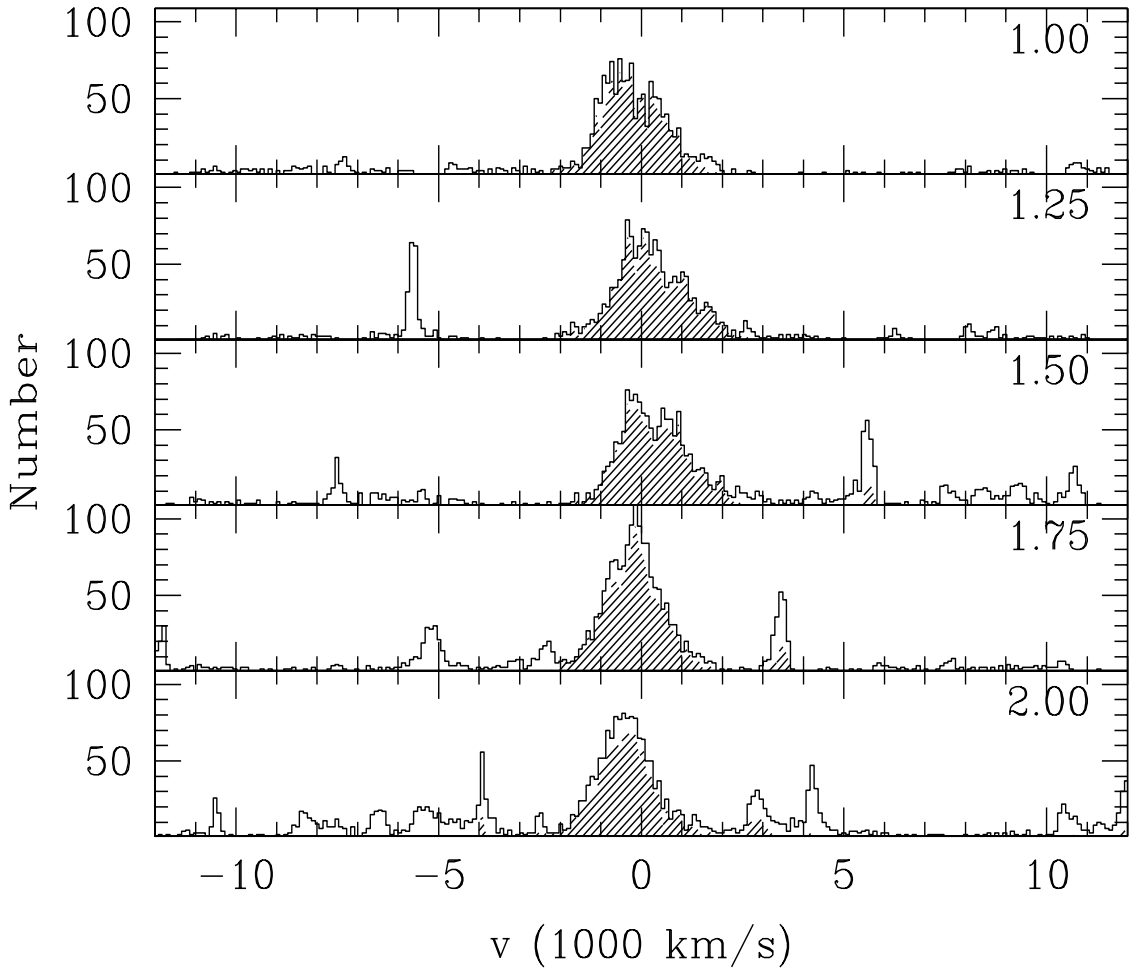}
\end{center}
\figcaption{Line of sight velocity histograms for the five lines of sight
through Cluster 6 previously examined in Fig.~\ref{fig:zetacl06z0.5};
numbers in the upper right--hand corner indicate the mass ratios
associated with each line of sight, for the simple projected estimator.
Material used in forming the histograms was taken from a viewing
``cylinder'' of radius $3 h^{-1}$Mpc; the space between tick marks
in the plot corresponds to $1000\ {\rm km/s}\,=\,10h^{-1}$Mpc, and
thus a volume associated with each tick mark of $283h^{-3}$Mpc$^3$.
The unshaded outline shows the histogram produced by the mass in
the line of sight, while the shaded subset of the histogram is
produced by mass at density contrasts above 50.
\label{fig:losvels}
}
\end{figure*}
at the high end of the mass function argued by
e.g. Bahcall \& Fan (\cite{BF}).

\section{Discussion}

\subsection{Detecting projection effects with line of sight velocities}
\label{ssec:losvels}

One way in which an observer might hope to avoid being fooled by this
sort of projection effect
is by examining line of sight velocities of galaxies.  If a velocity
histogram suggests a clump of mass in the foreground or background of
the cluster, an observer could assume that a lensing mass measurement
from this cluster would be corrupted.  While this would not remove the
effect on mass measurements of this cluster, it would remove the
cluster from consideration, and thus avoid any false scientific
conclusions drawn from an incorrect mass assigned to this cluster.
For instance, ignoring a mass estimate from a cluster which clearly
has a strong foreground or background mass concentration would then
reduce the chance of overestimating the abundance of high--mass clusters.

In constructing the histograms shown earlier, we assumed that an
observer would be perfectly able to avoid such lines of sight when
we explicitly excluded lines of sight which pass through another
large cluster in the volume.  Thus, the histograms at present
have attempted to correct for foreground and background contamination
by other objects.  However, it is still worth considering whether
galaxy redshifts could help us to exclude cases
where no big objects such as rich clusters ruin our measurement,
but instead a filament is oriented near or across the line of sight.

The simulations used for this project do not model galaxy formation.
Indeed, in the lowest resolution region (outside the cluster environs,
and comprising most of the volume of the simulation), individual
collisionless particles have masses
$\sim 6.66\times\ 10^{11} h^{-1} M_{\odot}$,
so approximating galaxies by halos is not available to us.  Instead,
we tie galaxy locations to the background overdensity field, and
consider two scenarios:  a simple model where galaxies trace the
mass along the line of sight; and a model where galaxies are biased
tracers of the mass, lying in regions where the mean local density
contrast is above 50.

Fig.~\ref{fig:losvels} shows the results of this process.  Here we
have taken five lines of sight through Cluster 6, corresponding
to $M_{\rm lens}/M_{\rm true}$ values of $1.0$ -- $2.0$ which were not
excluded in making the cluster histogram as no large lumps exist in
the cluster foreground or background.  These correspond to the same
lines of sight used in Fig.~\ref{fig:zetacl06z0.5}.  For each,
a line--of--sight
velocity histogram was constructed by considering particles taken
from the overall mass distribution and its velocity field.  Shown
within that histogram is a smaller shaded one, detailing the histogram
produced by matter at density contrasts of 50 or higher.

The velocity histograms determined by the full mass distribution
shows that if galaxies trace the mass, one cannot generically count
on line--of--sight velocities to veto cases with large errors,
even when densely sampled.  The last example shown here --- a
line of sight towards Cluster 6 producing an error in the mass
estimate of a factor of 2 --- is driven by comparatively diffuse
mass extending over a range of radii in the foreground:  a filament
of mass.  It is
not at all clear that a histogram of galaxy velocities would pick
this up.  Note in the line of sight with mass ratio 1.25 shown
here, a mass concentration 50--60$h^{-1}$Mpc in the foreground
corresponds to a filament cutting {\em across} our line of sight,
rather than to any one specific collapsed object.

The situation is even worse if galaxies are expected to
lie preferentially in regions of high overdensity.  As noted, the
shaded area indicates a histogram drawn from regions with overdensity
greater than 50; the difference between this histogram and the
parent (unshaded) histogram highlights the difference between the
expected sort of histograms should galaxies trace the mass
directly and should galaxies be biased towards higher--density
regions.  In the latter case, it is quite clear that clumps of
galaxies in a line of sight velocity diagram cannot be
counted upon to screen out any potentially large errors.

\subsection{Scale dependence of results}
\label{ssec:scaledep}

In studying the magnitude of this effect, we have been considering
matter contained within spheres of radius $128 h^{-1}$Mpc centered
on the cluster of interest.  Of course, in the real universe, when
observing a cluster along a particular line of sight, there exists
material at distances greater than $128 h^{-1}$Mpc away from the
cluster --- material in the foreground, between us and the cluster
lens, and in the background between the lens and the source, but
further away from the cluster than $128 h^{-1}$Mpc.  What effect
should this material have on the estimated lensing mass?

\begin{figure}[t]
\begin{center}
\leavevmode
\epsfysize=8cm \epsfxsize=8cm\epsfbox{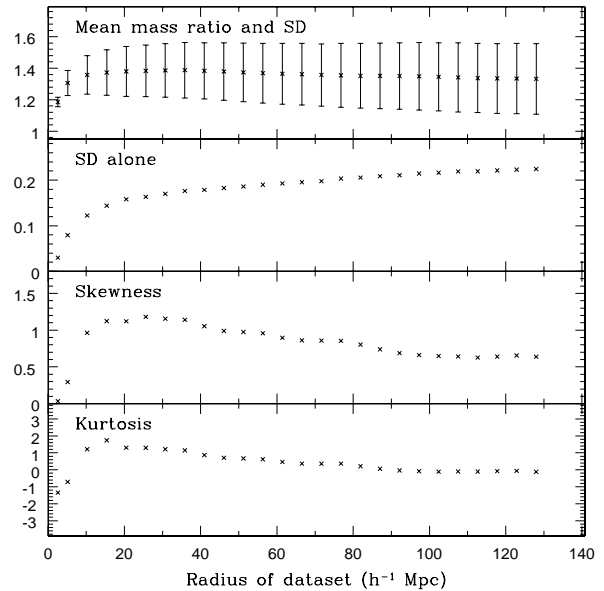}
\end{center}
\figcaption{Moments of the distribution of $M_{\rm lens}/M_{\rm true}$ for Cluster 6,
as a function of the outer radius of the particle dataset used to calculate
the distribution.  The data points showing the moments for the largest radius
corresponds to using the full simulation dataset, and thus are the moments
of the histogram for Cluster 6 shown in Fig.~\ref{fig:hist200z0.5}.
\label{fig:cl6moments}
}
\end{figure}

To examine this, we repeated the procedure used in constructing
Fig.~\ref{fig:hist200z0.5} for Cluster 6 several times; however,
instead of using the full simulation dataset (a sphere of radius
$128 h^{-1}$Mpc centered on the cluster), we used datasets centered
on the cluster but cut off at a different radius each time.
This allows study of the evolution of the histogram as a function
of the volume of matter around the cluster being considered in
calculating the effect.  In particular, we considered the first
four moments of the histogram for Cluster 6 at $z_{l}\,=\,0.5$;
the results of this analysis are shown in Fig.~\ref{fig:cl6moments}.
Each radial point corresponds to taking a sphere of material of
that radius, centered on the cluster, and using 5000 lines of
sight through that sphere to determine the moments of the distribution
of $M_{\rm lens}/M_{\rm true}$.  The last data points for each moment, at a
radius of $128 h^{-1}$Mpc (half the box size), corresponds to the
histogram for Cluster 6 shown in Fig.~\ref{fig:hist200z0.5}.
The most striking feature to note is that the mean value of the
histogram is reached at comparatively small radii for the dataset.
The mean error is driven primarily by material within 10 -- 20 Mpc
of the cluster.  As the radius of the dataset used decreases, the
histogram converges towards that driven by the asymmetry of the
cluster itself.  Also, the {\em shape} of the histogram appears to have
converged; the skewness and kurtosis of the histogram for material
with $100h^{-1}$Mpc remains relatively unchanged after considering
the effects of material at still larger radii.

The standard deviation of the histogram, on the other hand, appears
to still be increasing as the effect of material further and further
away from the cluster is considered.
One plausible interpretation of this result is as follows.
When only the material within a sphere of a given radius is considered,
a line of sight through the cluster will produce some lensing mass
estimate.  Considering a slightly larger sphere of material 
will slightly change the estimate produced along each line of sight,
by an amount dependent upon the overdensity contained within the
small additional segment of volume that lies along the line of sight.
Thus, the new histogram of this larger sphere can be thought of as
the histogram describing the smaller sphere {\em convolved} with a
function dependent upon the distribution of overdensities expected
in these volumes.  Therefore, at a large enough
radius from the cluster, this component should simply be noise
uncorrelated with the cluster itself; the result is to broaden
the histogram, resulting in an increasing second moment.  This
was checked by plotting the mass estimates arising from the
material within a given large radius (e.g. $123 h^{-1}$Mpc, the
second--to--last points in radius on Fig.~\ref{fig:cl6moments})
with the {\em change} in the mass estimate when all the material
(out to $128 h^{-1}$Mpc, the last point in Fig.~\ref{fig:cl6moments})
is used.  The result was no appreciable correlation, with a
correlation coefficient of $r\,=\,0.10$.

\subsection{Comparison With Other Work}

Papers by Cen (\cite{Cen}), Reblinsky \& Bartelmann (\cite{RebBart})
and Brainerd {\em et.al.} (\cite{BWGV}) have
previously examined the accuracy of weak lensing mass estimates.

Cen (\cite{Cen}) constructed
mock surface density maps by simply projecting down the mass in the
simulation, subtracting off a constant background since matter at
the background density does not contribute to the lensing signal.
In the limit of no variation of the lensing kernel in
Eq.~(\ref{eq:kappaint}), these approaches should be equivalent.
Cen reported substantially smaller positive bias in the lensing mass
than we have seen here, typically of only $5$--$10\%$.  On the
surface, this appears to be discrepant with our result.

However, masses compared by Cen were measured within an aperture
of fixed size, rather than an aperture whose size depends on the
mass within, as we use here.  Furthermore, the magnitude of the
bias seen by Cen depended strongly on the size of the aperture used,
and increased dramatically between apertures of radii $1 h^{-1}$Mpc
(a median value of just under 14\%, by eye from Fig. 19) and
$2 h^{-1}$Mpc (approximately 40\%).  Cen measured the bias and
dispersion of $M_{lens}/M_{200}$ using an aperture for each cluster
equal to its three--dimensional value of $r_{200}$, and found
results substantially similar to those for a fixed aperture of
$1 h^{-1}$Mpc.  This suggests that $1 h^{-1}$Mpc is representative
of the $r_{200}$ values for the 50 clusters in Cen's sample, which
was taken at $z\,=\,0$.
The 12 clusters in our sample represent the high--mass tail
of the cluster mass function, in a box 8 times as large;
the values of $r_{200}$ represented at $z\,=\,0.5$ range from
$1.01 h^{-1}$Mpc to $1.54 h^{-1}$Mpc.  The median value for the
bias we obtain using a fixed aperture for each cluster equal to that
cluster's 3D value of $r_{200}$, similar to Cen, is $24\%$;
the median using a fixed aperture of $2 h^{-1}$Mpc is 1.56.
We believe the difference in the magnitude of the bias seen
by Cen and ourselves lies in the different cluster samples used ---
Cen's sample running much further down the mass function ---
and also in sample variance.  With regard to the latter, the
medians for the mass ratios obtained from each cluster independently,
using a $2 h^{-1}$Mpc fixed aperture, range from $1.38$ to $1.85$.

We emphasize again that the dispersion in measured values of
the mass is the important quantity from Fig.~\ref{fig:pdfz0.5},
since any attempt by an estimator to correct for any bias from
projected material must still contend with the variance in
the amount of projected material along different lines of
sight.  Fig. 19 of Cen (\cite{Cen}) suggests approximately 6\% of
his cluster lines of sight have mass ratios $M_{lens}/M_{200}$
measured within the 3D $r_{200}$ with values above 1.58; with
our smaller, larger mass sample, we see $10\%$ of our lines
of sight above this value.  For a fixed projection aperture
of $2 h^{-1}$Mpc, Cen sees 10\% of his cluster observations
yielding mass ratios in excess of approximately 2.0; we see
10\% lying above 2.14.  Within the limits of sample variance,
we do not believe our results on the dispersion are significantly
different from Cen.  In any case, both our work and Cen (\cite{Cen})
imply that large values of the mass ratio are not uncommon.

In Reblinsky \& Bartelmann (\cite{RebBart}), mock shear maps of
simulated clusters
were obtained by first projecting the mass into two--dimensions
to construct the convergence, then solving the Poisson equation
for the 2D lensing potential, Eq.~\ref{eq:poisson}, and finally
taking the appropriate derivatives to find the complex shear.
The tangential shear was then used to find the $\zeta$--parameter
using Eq.~\ref{eq:zetadefshear}.  Finally, the system of equations
relating $\zeta(r_1,r_2)$ (for a range of values of $r_1$ out to $r_2$)
and the appropriate averages of the convergence $\kappa$ (see
Eq.~\ref{eq:zetadef}) were solved for the mass profile assuming
no convergence in the outermost annulus.  These masses were then
compared to the 3D mass, using $r_{3D}\,=\,r2\,=\,1.8 h^{-1}$Mpc.
While on
the surface their results appear consistent with ours --- means,
medians and dispersions which are roughly comparable to those
presented here --- the comparison is in fact difficult to make
directly.  The mass range of the cluster sample used by
Reblinsky \& Bartelmann lies a factor of 4--10 lower than that
used here.  Their positive bias in $M_{lens}$ is driven by their
lowest-mass cluster; higher mass systems appear to evidence a
trend towards {\em underestimating} masses.  However, their setting
the effective convergence to be zero at $1.8 h^{-1}$Mpc can be
interpreted as estimating the typical foreground/background
mass contamination and subtracting that off; different mass estimators
can be expected to have different means, or degrees of bias.  The
important comparison to be made is the dispersion in their mass
ratios, $0.34$, which actually exceeds that in our sample and
appears relatively insensitive of the subsample taken.  The
apparent trend in their sample of decreasing bias with larger
mass systems is not seen in our sample, although it was seen in
our previous letter (Metzler et al.~\cite{MWLN}).  Given the
sample--dependence of this trend in our work, and that the
trend seems much less prominent in the dataset of Reblinsky \&
Bartelmann when restricted to the highest--mass clusters more
akin to our ensemble, the reality of this trend is unclear.

In Brainerd {\em et.al.} (\cite{BWGV}), shear maps were constructed by
explicit ray--tracing through the volume of interest.  Estimated
lensing mass profiles were then derived from the shear maps
by assuming that the simulated cluster was a singular isothermal
sphere, in which case the cluster velocity dispersion (and thus
the mass profile) is a simple function of the average shear over
an annulus.  Brainerd {\em et.al.} found quite good agreement between
their estimated lensing masses and the actual cluster masses --
the lensing masses typically being $5$--$10\%$ underestimates
({\em cf.} the outermost circle--points in their Fig. 9).
However, their shear maps were constructed using only the matter
on their highest--resolution grid within their clusters' true,
three--dimensional value of $r_{200}$; the effect of structure
at larger distances from the cluster is not considered at all.
In other words, Brainerd {\em et.al.} probed primarily the effect of
the anisotropic shape of the cluster itself, which we agree (as
noted above) has a small effect on mass estimates.  However,
their study is not sensitive to the effect of the large--scale
structure in which the cluster is embedded, examined here.

\bigskip
\section{Summary}

In this paper, we have examined how weak lensing mass estimates of
galaxy clusters are affected by the large--scale structure in which
the clusters are embedded.  We find that cluster masses are typically
overestimated, with mean errors of tens of percent for clusters at
redshifts $z\,\simeq\,0.5$, although the
exact value depends on the mass estimator chosen; any mass estimator
used on real observations should be calibrated through methods such
as described here.  These errors are likely worse for clusters at
higher redshift ($z\,\simeq\,1.0$).  We also note that as long as there
exists
a dispersion in observed masses possible for a given actual mass
(e.g. along different lines of sight through the cluster),
{\em even an unbiased
estimator} is likely to produce observed mass estimates which are
systematically higher than the actual virial masses of clusters;
this is
simply because there are more low--mass clusters which can be
erroneously
assigned a high mass than there are high--mass clusters which can be
erroneously
assigned a low mass.  The magnitude of this effect is crucially
dependent on the estimator used and should be considered carefully
when implementing a particular estimator.  Such approaches as
simple projection and aperture densitometry do not appear to
perform better than simple virial estimates based on cluster
X--ray temperatures alone, although temperature--based estimates
suffer from calibration uncertainties that would induce a bias
if unresolved.  The effects of large--scale structure examined
here are not typically resolved through the examination of galaxy
redshifts in the viewing field.

\bigskip
The authors would like to thank Chris Kochanek and Ludovic van Waerbeke
for useful discussions,
Mike Norman for his assistance in using the LCA/MARG X--ray
Cluster Archive, and our referee for several valuable suggestions.
This research was supported by the National Science Foundation under
grant number PHY--0096151.

\end{document}